\documentclass[%
reprint,
amsmath,amssymb,
aps,
prb,
superscriptaddress,
showpacs, showkeys]{revtex4-1}

\usepackage{color}
\usepackage{graphicx}% Include figure files
\usepackage{dcolumn}% Align table columns on decimal point
\usepackage{bm}% bold math
\usepackage{hyperref}% add hypertext capabilities
\hypersetup{bookmarksnumbered, pdfpagemode=UseOutlines, 
colorlinks=true, citecolor=blue, filecolor=blue, linkcolor=blue, urlcolor=blue}

\usepackage{xcolor}	% \textcolor{<color>}{<text>}
\usepackage{verbatim} % for multiline comments
\graphicspath{C:\Users\p269017\Dropbox\Data_M_Gurram\FND\Talks_Wed-Sci_Tue-Group_FND\Group_Talk_II_Mallik_2015.02.17\2015_Manuscript
}

\begin{document}

%\preprint{APS/123-QED}

\title{Spin transport in fully hexagonal boron nitride encapsulated graphene}

\author{M. Gurram}
\thanks{corresponding author}
\email{m.gurram@rug.nl}
\affiliation{Physics of Nanodevices, Zernike Institute for Advanced Materials, University of Groningen, The Netherlands}
\author{S. Omar}
\affiliation{Physics of Nanodevices, Zernike Institute for Advanced Materials, University of Groningen, The Netherlands}
\author{S. Zihlmann}
\affiliation{Department of Physics, University of Basel, Basel, Switzerland}
\author{P. Makk}
\affiliation{Department of Physics, University of Basel, Basel, Switzerland}
\author{C. Sch\"{o}nenberger}
\affiliation{Department of Physics, University of Basel, Basel, Switzerland}
\author{B.J. van Wees}
\affiliation{Physics of Nanodevices, Zernike Institute for Advanced Materials, University of Groningen, The Netherlands}%

\date{\today}% It is always \today, today,

\begin{abstract}
We study fully hexagonal boron nitride (hBN)-encapsulated graphene spin valve devices at room temperature. The device consists of a graphene channel encapsulated between two crystalline hBN flakes; thick-hBN flake as a bottom gate dielectric substrate which masks the charge impurities from SiO$_2$/Si substrate and single-layer thin-hBN flake as a tunnel barrier. Full encapsulation prevents the graphene from coming in contact with any polymer/chemical during the lithography and thus gives homogeneous charge and spin transport properties across different regions of the encapsulated graphene. Further, even with the multiple electrodes in between the injection and the detection electrodes which are in conductivity mismatch regime, we observe spin transport over 12.5 $\mu$m long distance under the thin-hBN encapsulated graphene channel, demonstrating the clean interface and the pin-hole free nature of the thin-hBN as an efficient tunnel barrier.

% \begin{description}
% \verb+85.75.-d+, \verb+73.22.Pr+, \verb+75.76.+j+, \verb+73.40.Gk+
 % spin polarized transport devices, 85.75.-d
 % graphene, 73.22.Pr
 % spin transport effects, 75.76.+j %CHECK: this printed as j+ instead of +j
 % tunneling in interface structures, 73.40.Gk
% \end{description}
\end{abstract}

% \keywords{Spintronics, Graphene, Boron nitride, Tunnel barrier, Full encapsulation}%Use showkeys class option if keyword
\maketitle

%\tableofcontents
\section{Introduction}

Graphene is considered as an ideal material for spintronics due to its low intrinsic spin-orbit coupling, small hyperfine interaction, tunable carrier density and high electron mobility \cite{25.2009Neto_electronic_properties, 26.2011Sdas_etransport, 27.2011Dugaev_SOI} even at room temperature which is important for future applications. Theoretically it is estimated to show long spin relaxation length ($\lambda_{s}$) of 100 $\mu$m and high spin relaxation time ($\tau_{s}$) of 100 ns \cite{28.2006Macdonald_RSOI, 29.2012Macdonald_spintronics}. At the moment, experimentally these values are reported up to $\lambda_{s}$ = 30.5 $\mu$m and $\tau_{s}$ = 12.6 ns \cite{41.2016.Drogler_12ns}. In a quest to figure out the sources limiting the intrinsic spin transport in graphene, there have been several experiments, which suggest that the role of underlying substrate and the quality of tunnel barrier is crucial\citep{39.2014.Stephan_challenges}.

The quality of substrate plays an important role in determining the charge and the spin transport characteristics of graphene. An atomically flat hexagonal boron nitride (hBN) substrate has shown to provide a good base for high mobility graphene \cite{18.Dean_1Dcontacts} and long spin relaxation length\citep{14.Paul_GrhBN} because hBN has less charged impurities and the thickness of the bottom-hBN moves the graphene away from the charged impurities in the SiO$_{2}$ substrate. Also the effect of the electron-hole puddles can be different for different substrates and it is lower with a hBN substrate\cite{37.2015.Dinh_spin_ehpuddles}.

Recently, Kretinin \textit{et al}.\cite{10.Kretinin_hBNGrMoS2_NL} demonstrated the high electronic quality of graphene by encapsulating it from surroundings with different two-dimensional (2D) crystals such as hBN and transition metal dichalcogenide (TMD). However, this device geometry comprised of one dimensional edge electrodes \citep{18.Dean_1Dcontacts} with a thick semiconducting TMD top layer which is not compatible for standard four probe non-local spin transport measurements. In the present work, we replace the TMD with an atomically thin hBN flake, which serves two purposes. First, it acts as an encapsulating layer. Second, it acts as a tunnel barrier replacing the conventional oxide tunnel barrier. Therefore it allows to probe the charge and the spin transport properties of the encapsulated graphene with multiple ferromagnetic (FM) electrodes.

The previous studies have shown that the graphene interface with spin sensitive ferromagnetic electrodes affects the spin transport behaviour. Further, the conventional oxide tunnel barriers used in spintronic devices suffer from pinholes, inhomogeneous coverage, non-uniform growth, and consequently contributing to the low spin transport properties\cite{30.2014Aachen_contacts_oxidation, 39.2014.Stephan_challenges}. A 2D-layer of thin-hBN with close lattice match to graphene has been shown to exhibit pin-hole free tunnelling characteristics \cite{38.2011.Lee_hBNtunnel, 6.Britnell_hBNtb}. It is also predicted to enhance the spin injection efficiency from a ferromagnet into graphene \cite{7.Theory_hBNtb}.

Yamaguchi \textit{et al}. \cite{2.Yamaguchi_exhBNtb} first reported spin injection through exfoliated monolayer hBN tunnel barrier into bilayer graphene resulting in quite low spin transport parameters ($\tau_s$ = 56 ps, D$_s$ = 0.034 m$^2$/s, $\lambda_s$ = 1.35 $\mu$m) and low mobility ($\mu$ = 2700 cm$^2$/Vs) of graphene. Further, Kamalakar \textit{et al}.\cite{3.Kamalakar_CVDhBNtb} and Fu \textit{et al}.\cite{4.Basel_CVDhBNtb} used large scale chemical vapour deposited (CVD) bilayer hBN-tunnel barriers for spin injection into graphene. The spin polarization increased with CVD hBN barriers. However, the quality of graphene channel is limited by the wet transfer technique used for transferring CVD hBN on to graphene. In contrast, we use a dry pick-up and transfer technique\cite{19.Paul_fastpickup} in order to completely cover the underlying graphene with a thin-hBN flake, which acts a tunnel barrier and also shields the graphene from external polymers and chemicals during the lithography process.

It is also worth discussing here that the previous reports on spin transport in hBN encapsulated graphene are quite promising in terms of improved spin transport parameters\cite{17.Marcos_hBNencapsulation, 16.Pep_24um}. However extracting the correct spin transport parameters is not straightforward due to inhomogeneous spin transport behaviour of the hBN-encapsulated and the non-hBN-encapsulated graphene regions. A similar behaviour is also observed for partly suspended high mobility graphene \cite{15.Marcos_GrSuspended}. Using our device geometry, we can achieve more homogeneous charge and spin transport behaviour in the graphene channel compared to the previously reported results.

In this work, we report a fully hBN-encapsulated graphene spintronics device to overcome the three aforementioned challenges, namely, 1) the influence of underlying substrate, 2) the influence of tunnel barrier interface, and 3) the inhomogeneity in graphene channel. We use a dry pick-up technique which prevents the graphene from external doping and results in more homogeneous charge and spin transport parameters at room temperature.

\section{device fabrication}
The hBN/graphene/hBN stack is prepared following the dry pick-up and transfer technique\cite{18.Dean_1Dcontacts, 19.Paul_fastpickup}. The graphene and hBN flakes were obtained by the exfoliation of highly oriented pyrolytic graphite (HOPG, SPI Supplies, ZYA grade) and hBN powder (HQ graphene). A 90 nm thick SiO$_{2}$/Si substrate is used for exfoliation and identification of graphene and thin-hBN flakes as it gives a good optical contrast to search for thinner graphene\cite{1b.Hunting_graphene} and hBN\cite{1_Hunting_hBN} flakes down to a monolayer. 

A polydimethylsiloxane (PDMS) polymer stamp prepared with polycarbonate (PC) layer is used to pick-up the flakes. At first, a glass substrate with PDMS/PC is used to pick-up a thick top-hBN flake which is used to pick-up the thin-hBN flake followed by picking up the graphene flake. Then the PC/top-hBN/thin-hBN/graphene stack is released onto a thick bottom-hBN on a SiO$_{2}$(300 nm)/Si substrate by melting the PC layer. Next, the PC layer is dissolved in chloroform for 5 hours at 50 $^{\circ}$C, followed by annealing in Ar/H$_{2}$ atmosphere at 350 $^{\circ}$C for 12 hours to remove the leftover PC residues on top of the thin-hBN and the top-hBN. It is important to note that the graphene beneath the thin-hBN does not come in contact with polymers during the fabrication due to the full encapsulation by hBN flakes.

\begin{figure}[!ht]
 \includegraphics[width=\columnwidth,trim= 0in 0in 0in 0in,clip]{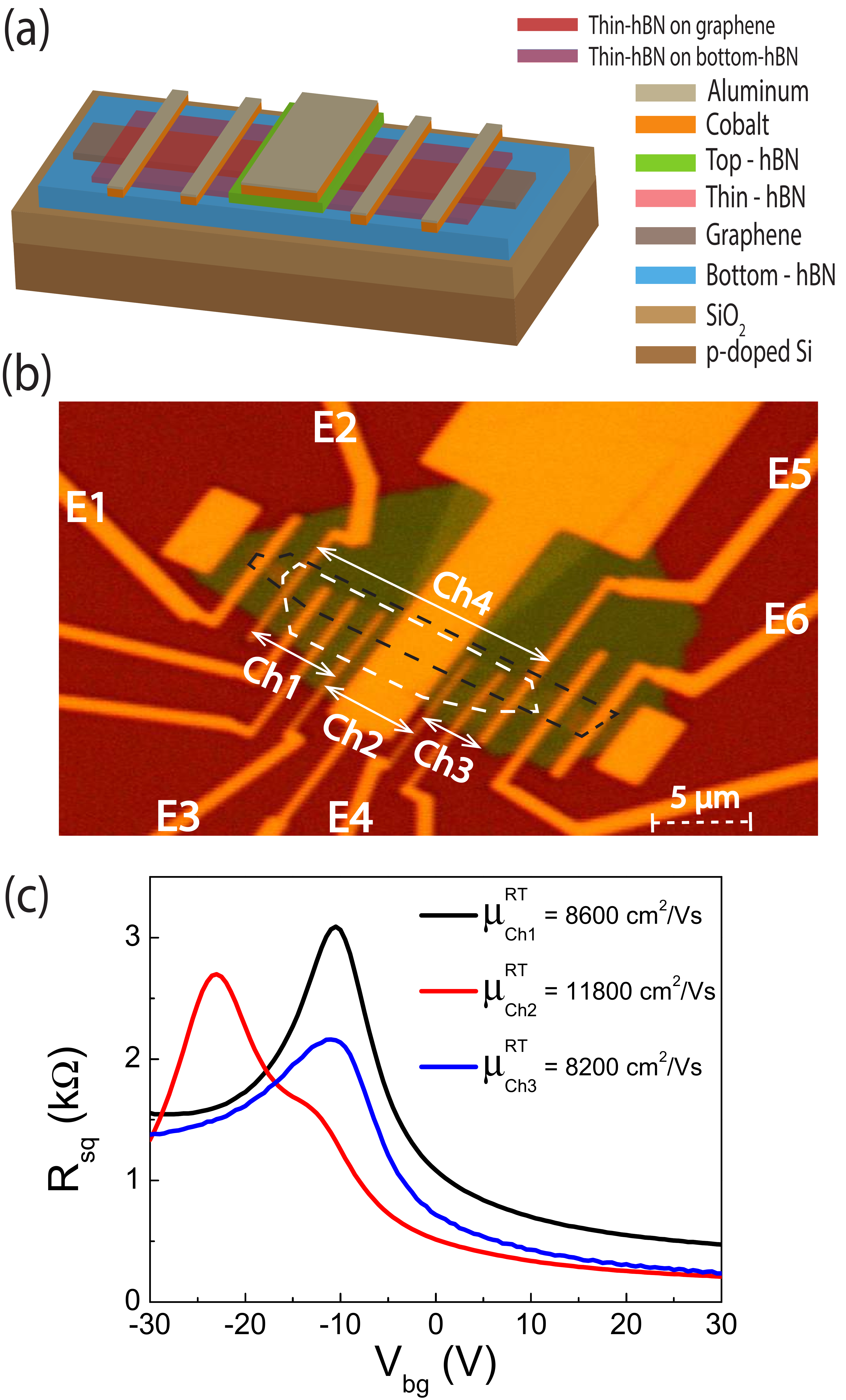}% Here is how to import EPS art
 \caption{\label{fig:Figure_1} (a) Schematic of the proposed fully hBN-encapsulated graphene spin valve device. (b) An optical microscope image of the fabricated device on SiO$_{2}$(300 nm)/Si substrate with multiple ferromagnetic cobalt electrodes (65 nm) with aluminum capping layer (4 nm) denoted as E(1-6). The graphene flake (0.4 nm) is completely supported underneath by the thick bottom-hBN (21 nm) flake and encapsulated by the thin-hBN (0.52 nm) layer between the electrodes E2 and E5. The different regions of the thin-hBN encapsulated graphene channel are denoted as Ch(1-4) with the double arrows. The boundaries of graphene and thin-hBN flakes are outlined with black and white dashed lines respectively. (c) Room temperature square resistance (R$_{sq}$) for different graphene channel regions as a function of backgate voltage (V$_{bg}$). The corresponding electron mobility values are given in the legend.}
\end{figure}

Thereafter, electron-beam lithography is used for patterning of electrodes on the poly(methyl methacrylate) (PMMA) coated stack followed by electron-beam evaporation to deposit 65 nm of ferromagnetic cobalt (Co) for spin sensitive electrodes. Cobalt is capped with 4 nm thick aluminum (Al) layer to prevent it from oxidation. A schematic of the layer-by-layer device structure is shown in Figure ~\ref{fig:Figure_1}(a). 

An optical micrograph of the fabricated device is shown in Figure ~\ref{fig:Figure_1}(b) where the graphene and the thin-hBN flakes are outlined with the black and the white dashed lines respectively. Due to a slight misalignment during the pick-up process, a region of 0.1 - 0.2 $\mu$m width along the top edge of the graphene flake is not covered by the thin-hBN. However, the electrodes in between E2 and E5 are deposited on top of the thin-hBN layer, avoiding the uncovered top-edge graphene region. The thin-hBN encapsulated graphene does not come in contact with polymers whereas the graphene regions which are not covered by the thin-hBN are touched by PC and PMMA polymers during the fabrication.
%The contacts C2 and C5 lie on the edges of thin-hBN, also partially on the graphene flake.

We performed atomic force microscopy (AFM) imaging to find the thickness of each flake used in the device fabrication. It is found to be 0.40 nm for graphene and 0.52 nm for thin-hBN layer. These values are in close agreement with the experimentally reported thickness for single layer (1L)-graphene \cite{23.Nikos_2007} and 1L-hBN \cite{3.Kamalakar_CVDhBNtb, 6.Britnell_hBNtb}.

\section{Results and Discussion}

We report the measurements for the different regions of fully encapsulated graphene channel as labelled in the optical image shown in Figure ~\ref{fig:Figure_1}(b). Specifically, Ch1 (4.5 $\mu$m, between E2 and E3) and Ch3 (3 $\mu$m, between E4 and E5) regions consist of graphene encapsulated by thin-hBN. Ch2 (5 $\mu$m, between E3 and E4) region comprised of graphene encapsulated by the thin-hBN. On top of the thin-hBN, we put thick top-hBN (6nm) flake which serves as a topgate dielectric. Finally, Ch4 (12.5 $\mu$m, between E2 and E5) consists of region across the whole thin-hBN encapsulated graphene.

The resistance area product (R$_{c}$A) for graphene/hBN/graphene tunnel junction is reported to scale exponentially with the number of hBN layers \cite{6.Britnell_hBNtb}. We characterize the thin-hBN/graphene interface resistance (R$_{c}$) using a three probe measurement scheme. The R$_{c}$A product for the electrodes E(2-5) is found to be in the range of 0.3 - 1.1 k$\Omega \mu$m$^{2}$ which agrees with the reported values for single-layer hBN tunnel barriers \cite{6.Britnell_hBNtb,3.Kamalakar_CVDhBNtb, 4.Basel_CVDhBNtb}. Both the AFM and the R$_{c}$A analysis confirm the single-layer nature of the thin-hBN flake.

All the measurements are carried out at room temperature in a vacuum of $10^{-6}$ mbar. Charge transport measurements are carried out in a local four probe measurement scheme. An ac current is passed between electrodes E1 and E6 (Fig.~\ref{fig:Figure_1}(b)), and the voltage drop is measured across the electrodes for different transport channel regions in between. The measured square resistance (R$_{sq}$) is presented in Figure ~\ref{fig:Figure_1}(c) for three different graphene channel regions Ch1, Ch2, and Ch3.
The square resistance for the thin-hBN encapsulated regions (in Ch(1-3)) show charge neutrality point (CNP) around -12 V and for the top-hBN encapsulated part of the Ch2 region show it around -23 V. A small bump at -12 V in R$_{sq}$ data for Ch2 region corresponds to the thin-hBN encapsulated parts on either side of the top-hBN. 

The field effect mobility ($\mu$) of graphene channel is calculated by fitting the R$_{sq}$ data (Fig.~\ref{fig:Figure_1}(c)) using the relation, R$_{sq}$ = 1/(ne$\mu$ + $\sigma_{0}$) + $\rho_{s}$, where e is the electronic charge, n is the carrier density, $\sigma_{0}$ is the residual conductivity at the CNP, and $\rho_{s}$ is the contribution from short range scattering \cite{14b.Morozov_mobility, 14c.Paul_mobility}. The fitted values of the electron mobility for three different regions are $\mu_{Ch1}$ = 8600 cm$^{2}$/Vs, $\mu_{Ch2}$ = 11800 cm$^{2}$/Vs and $\mu_{Ch3}$ = 8200 cm$^{2}$/Vs. The close values of mobility at different regions reflect that the graphene channel is homogeneous under the thin-hBN encapsulation. A relatively higher value of mobility for Ch2 region is attributed to the central top-hBN encapsulated graphene while the remaining thin-hBN encapsulated graphene (Ch1 and Ch3) on either side of top-hBN has an uncovered graphene edge of 0.1 - 0.2 $\mu$m width which are exposed to polymers during the fabrication.

\begin{figure}
 \includegraphics[width=\columnwidth,trim= 0in 0in 0in 0in,clip]{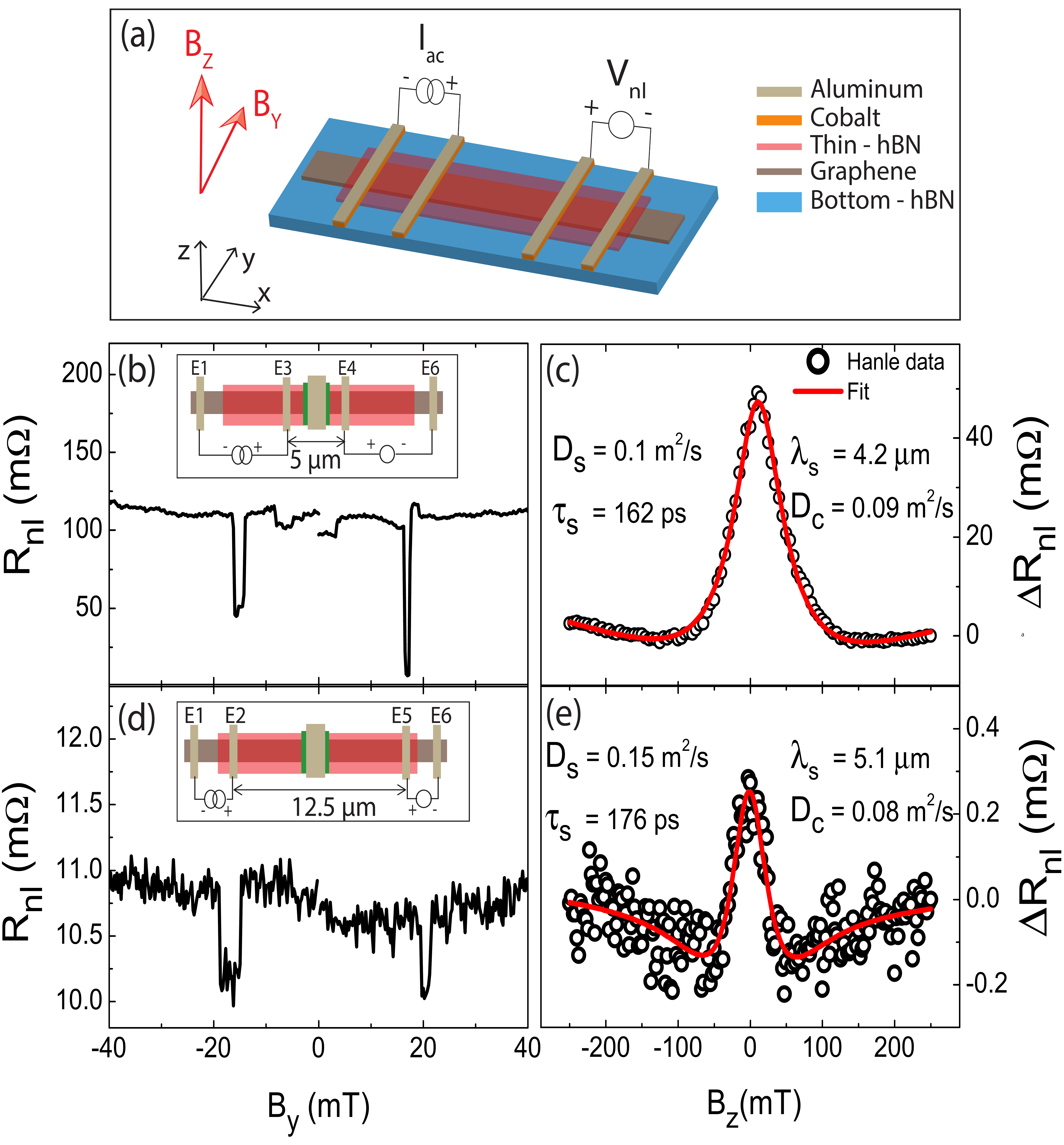}% Here is how to import EPS art
 \caption{\label{fig:Figure_2} (a) Schematic of the four probe non-local geometry for the spin valve and the Hanle measurements. Spin valve signals measured at V$_{bg}$ = 30 V for the top-hBN encapsulated (Ch2, between electrodes E3 and E4) region (b), and across the whole thin-hBN encapsulated (Ch4, between electrodes E2 and E5) region (d). The corresponding Hanle signals and fitting curves (c) and (e). The insets in (b) and (d) show the measurement geometry and the length of the graphene channel.}
\end{figure}

% Spin transport measurements are performed in a four probe non-local geometry as shown in Figure ~\ref{fig:Figure_2}(a). 
The spin transport measurements are performed in a four probe non-local geometry (Fig.~\ref{fig:Figure_2}(a)) at room temperature using standard ac lock-in technique with currents of 10 - 20 $\mu$A. We inject a spin polarized current using an ac current source (I$_{ac}$) and measure the non-local voltage (V$_{nl}$) while sweeping an in-plane magnetic field (B$_{y}$) parallel to the long axis of the ferromagnetic electrodes. The width of the electrodes is varied from 0.2 $\mu$m to 0.8 $\mu$m in order to switch magnetization of the electrodes at different coercive fields. As the B$_{y}$ is swept from a negative magnetic field to a positive field, steps in R$_{nl}$(=V$_{nl}$/I$_{ac}$) are observed whenever the magnetization of the two inner electrodes changes between parallel and anti-parallel configurations. The influence of the outer electrodes will be diminished if chosen far from the injection/detection electrodes, resulting in a typical two-level spin valve signal. In order to asses the spin transport nature of graphene, we extract the spin relaxation parameters from the Hanle spin precession measurements.

Hanle measurements are carried out in a four probe non-local spin valve geometry as shown in Figure  ~\ref{fig:Figure_2}(a) while the magnetic field is applied perpendicular (B$_{z}$) to the device plane. As the polarization of the injected spins is along y-direction, B$_{z}$ causes the spins to precess in x-y plane with a Larmor frequency $\omega_{L}$ = $g_e\mu_B$B$_{z}$/$\hbar$, where the Lande g factor g = 2, $\mu_{B}$ the Bohr magneton and $\hbar$ the reduced Plank constant. In order to eliminate the background magnetoresistance effects, we analyze the effective Hanle spin signal, $\Delta$R$_{nl}$ = (R$_{nl}^{\uparrow\uparrow}$ - R$_{nl}^{\uparrow\downarrow}$)/2, where R$_{nl}^{\uparrow\uparrow(\uparrow\downarrow)}$ is the Hanle signal for the parallel(anti-parallel) magnetization configuration of the injection and the detection electrodes. The resulting precession data $\Delta$R$_{nl}$ is fitted with the stationary solutions to the 1D Bloch equation in the diffusion regime; D$_{s}\bigtriangledown^{2}\vec{\mu_{s}} - \vec{\mu_{s}}/\tau_{s} + \vec{\omega_{L}}\times\vec{\mu_{s}}= 0$, where $\mu_{s} $ is the spin chemical potential, D$_{s}$ is the spin diffusion constant, and $\tau_{s}$ is the spin relaxation time. 

The spin valve signals for the Ch2 region and across the whole thin-hBN encapsulated (Ch4) region are shown in Figure ~\ref{fig:Figure_2}(b) and ~\ref{fig:Figure_2}(d), measured at the backgate voltage of 30 V. The corresponding Hanle signals and their fitting curves are shown in Figure ~\ref{fig:Figure_2}(c) and ~\ref{fig:Figure_2}(e). 
%The spin relaxation parameters $\tau_{s}$ and D$_{s}$ are obtained from fitting the Hanle data with 1D Bloch equation. 
%A similar spin valve and Hanle signals are also observed for Ch1 and Ch3 transport channels. 

The extracted value of $\tau_{s}$ for different channel regions (Ch(1-4)) is in the range of 135 - 176 ps, D$_{s}$ is of 0.11 - 0.18 m$^{2}$/s, and the corresponding value of the spin relaxation length $\lambda_{s}$ (= $\sqrt{D_{s}\tau_{s}}$) is of 4.2 - 5.1 $\mu$m.  
The charge diffusion coefficient (D$_{c}$) calculated from the resistivity (R$_{sq}$) data in Figure ~\ref{fig:Figure_1}(c) is in the range of 0.07 - 0.09 m$^{2}$/s. As the values of D$_{c}$ and D$_{s}$ are in a reasonable agreement, we confirm the reliability of Hanle fitting \cite{34.2012Thomas_mismatch, 15.Marcos_GrSuspended}.

As we can see from the Hanle fitting data (Fig. ~\ref{fig:Figure_2}(c) and ~\ref{fig:Figure_2}(e)), the spin relaxation parameters do not differ much for different encapsulated regions under thin-hBN. Besides, the mobility values also lie close to each other (Fig. ~\ref{fig:Figure_1}(c)). It indicates that a consistent charge and spin transport behaviour is observed across different regions of the thin-hBN encapsulated graphene.
%the mobility values lie close to each other
%spin relaxation parameters are in close agreement for different encapsulated regions under thin-hBN.

The values of spin relaxation parameters are quite low compared to the graphene on hBN \citep{14.Paul_GrhBN} or even partly encapsulated by hBN \citep{17.Marcos_hBNencapsulation}. The thin-hBN flake is of single-layer and it resulted in low interface resistance (R$_c$) values for the electrodes E(2-5), 0.6 - 2.1 k$\Omega$ which lie in the same order of the spin resistance (R$_\lambda$) for the thin-hBN encapsulated graphene, 0.4 - 1.4 k$\Omega$ where R$_\lambda$ = R$_{sq}\lambda_s$/W, W = 1.8 $\mu$m is the width of the graphene. These values imply that the device is within the conductivity mismatch regime \citep{34.2012Thomas_mismatch}. A similar behaviour is also reported by Yamaguchi \textit{et al}.\cite{2.Yamaguchi_exhBNtb} with exfoliated 1L-hBN and Fu \textit{et al}.\cite{4.Basel_CVDhBNtb} with CVD 1L-hBN tunnel barriers.

It is also important to note in our device that the ferromagnetic cobalt electrodes on top of the thin-hBN layer are deposited in order to avoid contact with the uncovered (by thin-hBN) graphene edge (Fig. ~\ref{fig:Figure_1}(b)). The proximity of the stray magnetic field from the ends of the cobalt electrodes (between E2 and E5) can act as an additional dephasing field and influence the spin transport in graphene. Also the uncovered region is exposed to the PC and PMMA polymers during the fabrication which can reduce the mobility and spin relaxation time.

Further, we would like to emphasize that the spin diffusion is detected across 12.5 $\mu$m length of thin-hBN encapsulated (Ch4) graphene region (Fig.~\ref{fig:Figure_2}(d-e)) which consists of multiple electrodes in between the spin injection (E2) and the spin detection (E5) electrodes (Fig. ~\ref{fig:Figure_1}(b)). The conventional oxide-barrier/graphene interfaces are reported to act as spin sinks especially when their R$_c$ lie close to R$_\lambda$\cite{31.2010Kawakami_3tunnelbarriers, 30.2014Aachen_contacts_oxidation, 34.2012Thomas_mismatch} which hinders the long distance spin transport in graphene with multiple electrodes. Whereas, within the conductivity mismatched regime, our device still performs better than the device with oxide barriers, which might have pin-holes. Hence, we attribute the observed long distance spin transport behaviour to the pin-hole free nature of the thin-hBN layer and to the clean interface of thin-hBN with the graphene compared to the deposited oxide tunnel barrier.

We would also like to point out that we do not observe the very long spin relaxation length or high spin relaxation time as in the case of partly hBN-encapsulated graphene stack reported by Guimar\~{a}es \textit{et al}. \cite{17.Marcos_hBNencapsulation} and it is worthwhile to mention the differences between our device and the stack reported in ref. \onlinecite{17.Marcos_hBNencapsulation}.  In our device, PMMA residues lie at the thin-hBN/cobalt electrode interface which might affect the spin injection/detection in a different way than in the stack reported in ref. \onlinecite{17.Marcos_hBNencapsulation}, which has PMMA residues between the graphene and the oxide tunnel barrier. Further, our device shows lower mobility and charge diffusion constant which also translates to the lower spin relaxation length.

\begin{figure}[!ht]
 \includegraphics[width=\columnwidth,trim= 0in 0in 0in 0in,clip]{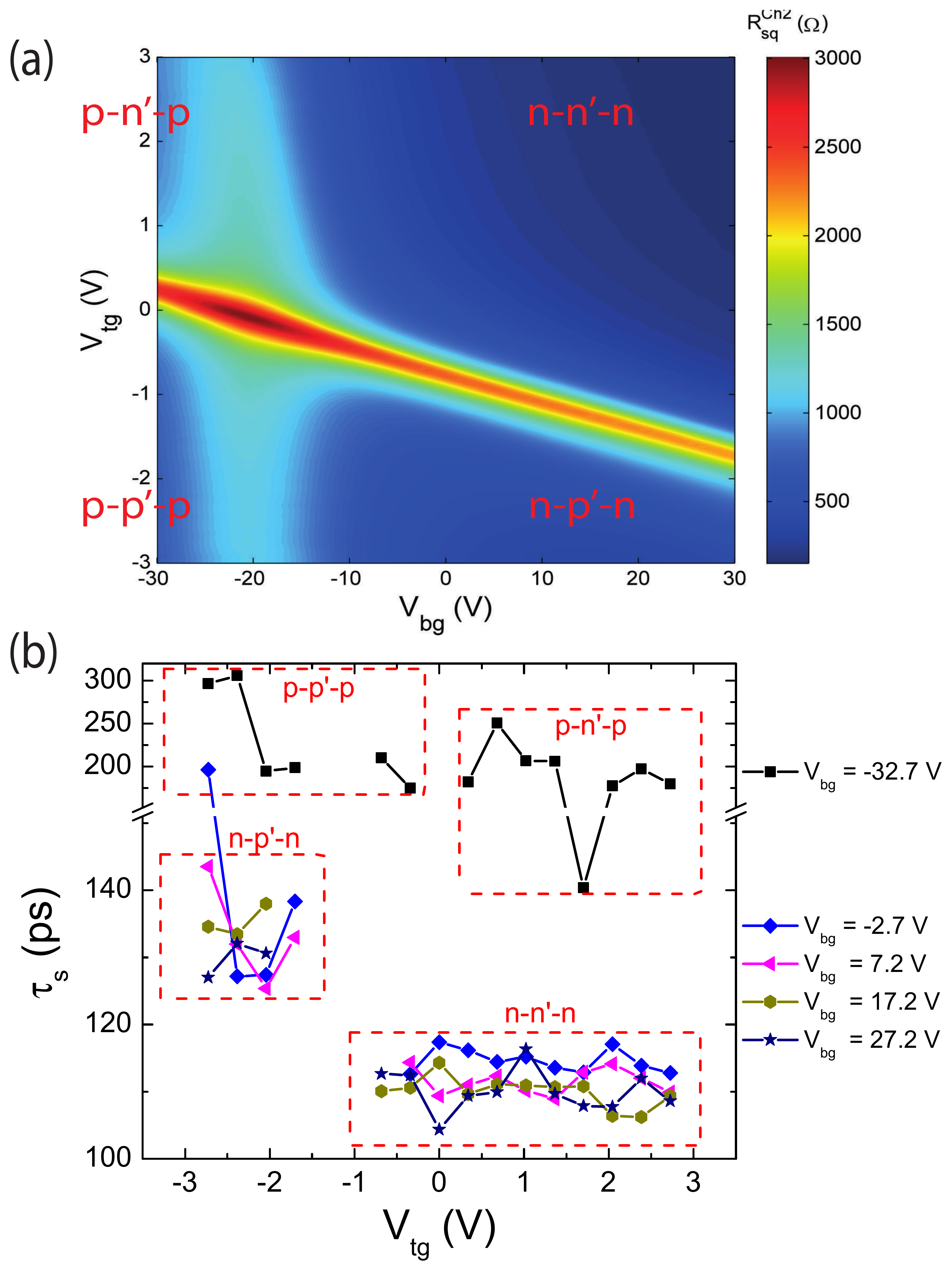}% Here is how to import EPS art
 \caption{\label{fig:Figure_3} (a) Contour plot of square resistance, R$_{sq}$ for top-hBN encapsulated (Ch2) region measured at room temperature with respect to backgate (V$_{bg}$) and topgate (V$_{tg}$) voltages. The polarities, p or n (p' or n') type of non-topgated (topgated) encapsulated regions are tuned independently using the combinations of V$_{bg}$ and V$_{tg}$ which resulted in n-n'-n, n-p'-n, p-n'-p and p-p'-p junction configurations. (b) Spin relaxation time as a function of topgate voltage at different backgate voltages. The dashed red lines outline the data corresponding to the four junction configurations.}
\end{figure}

Furthermore, using our device structure we can control the electric field and the carrier density independently with the top and the bottom gate electrodes. This allows us to study i) the electrical control of spin information in graphene\cite{17.Marcos_hBNencapsulation}, and ii) the spin transport across the p/n junctions created by the topgate and the non-topgate encapsulated graphene. It is interesting to study the spin transport across the p/n junction, which acts as a barrier for the transmission of spins, and results in high magnetoresistance and sensitivity in a spin valve transistor\cite{40.2005.Huang_SV_pn_transistor}.

A contour plot for the square resistance of the Ch2 region (R$_{sq}^{Ch2}$) as a function of V$_{tg}$ and V$_{bg}$ is shown in Figure ~\ref{fig:Figure_3}(a). The charge neutrality point (CNP) appears as a line with a slope of -0.034 which is equivalent to the ratio of the top and bottom gate dielectric capacitances. The maximum R$_{sq}$ of the CNP appears at V$_{bg}$ $\approx$ -22.7 V, V$_{tg}$ $\approx$ 0 V and decreases along either side of the line. It is a characteristic behaviour observed with single layer graphene \cite{17.Marcos_hBNencapsulation}. The V$_{tg}$ independent feature around V$_{bg}$ = -22.7 V appears due to the sides of the top-hBN encapsulated region that are non-topgated between the electrodes E3 and E4 \cite{16.Pep_24um}.

We can create an effective electric field in the top-hBN encapsulated part of the Ch2 region by modulating the topgate voltage (V$_{tg}$) and the backgate voltage (V$_{bg}$)\cite{35.2009Zhang_VbgVtgelectricfield}. The perpendicular electric field can induce a Rashba spin orbit field in graphene\cite{17.Marcos_hBNencapsulation} which can be used to manipulate the spin transport properties of the topgate encapsulated graphene channel.

The maximum value of the electric field created within the range of applied V$_{tg}$ and V$_{bg}$ is 0.22 V/nm. We do not observe a significant dependence of the spin relaxation time on the electric field. We attribute this to the lower field compared to the Rashba field (0.7 V/nm) applied in the case of Guimar\~{a}es \textit{et al}.\cite{17.Marcos_hBNencapsulation}.

Within the Ch2 region, carrier density in non-topgate encapsulated parts controlled by V$_{bg}$ whereas the central topgate encapsulated part is controlled by both V$_{bg}$ and V$_{tg}$. This clearly resulted in four quadrants representing different p/n junction configurations\citep{36.2007.Williams_pnjunctions} as indicated in Figure ~\ref{fig:Figure_3}(a). Due to the novelty of our device fabrication we could see four quadrants compared to the similar geometry reported in ref. \onlinecite{17.Marcos_hBNencapsulation} and \onlinecite{16.Pep_24um}. The non-topgated graphene regions in ref. \onlinecite{17.Marcos_hBNencapsulation} and \onlinecite{16.Pep_24um} are highly n-doped by the polycarbonate. However, in our case, graphene in the topgated region as well as the non-topgated region is protected by the thin-hBN and is not doped by polymers. So we are also able to access the p-doped characteristics of the non-topgated regions.

A strong dependence of the spin relaxation time ($\tau_s$) among the four junction configurations is observed as shown in Figure ~\ref{fig:Figure_3}(b). On average, the relaxation time is increased when the n(n')-doped region changes to p(p')-doped region. This can be observed at different topgate voltages as we move from n-n'-n region to n-p'-n region, and further to p-p'-p region. A possible explanation for the observed behaviour is that the contacts can induce a slight doping, as can be seen from the asymmetric square resistance of graphene in Figure ~\ref{fig:Figure_1}(c). This might result in an additional p-n interface in the p-doped regime, decoupling the contacts from the channel and resulting in higher spin relaxation time for holes. Furthermore, the gate dependence of spin lifetime in Figure ~\ref{fig:Figure_3}(b) does not show a dip around the CNP\cite{17.Marcos_hBNencapsulation, 21.Kawakami_1L2L} possibly due to the influence of p-n junctions formed at the edges of the topgated graphene channel near the CNP.

\section{conclusions}
In conclusion, for the first time we demonstrate a fully hBN-encapsulated graphene device for spintronics applications. We show that the full encapsulation of graphene results in homogeneous charge and spin transport properties at room temperature. Charge transport measurements show uniform mobility across different regions of the encapsulated graphene. Spin transport measurements show that a uniform spin relaxation length across different channel regions is achieved with the crystalline hBN encapsulation. Further, our device shows spin transport across the whole thin-hBN encapsulated region of 12.5 $\mu$m length even in the presence of conductivity mismatch electrodes, demonstrating the potential of using hBN as a tunnel barrier for graphene spintronics. The dual gate geometry allowed us to study the effect of electric field on the spin transport. However we do not observe a significant dependence due to the low values of achieved electric field. Moreover, we observe a strong dependence of the spin relaxation time on different p/n junction configurations. Further investigation is necessary to explain this behaviour.

We kindly acknowledge J.G. Holstein, H.M. de Roosz, H. Adema and T.J. Schouten for the technical assistance. We thank J. Ingla-Ayn\'{e}s for help in sample preparation and for the fruitful discussions. The research leading to these results has received funding from the European Union Seventh Framework Programme under grant agreement n$^{\circ}$604391 Graphene Flagship and supported by the Zernike Institute for Advanced Materials and the Netherlands Organization for Scientific Research (NWO).
  
% \bibliography{paper_mallik}	% Produces the bibliography via BibTeX.

\begin{thebibliography}{33}%
\makeatletter
\providecommand \@ifxundefined [1]{%
 \@ifx{#1\undefined}
}%
\providecommand \@ifnum [1]{%
 \ifnum #1\expandafter \@firstoftwo
 \else \expandafter \@secondoftwo
 \fi
}%
\providecommand \@ifx [1]{%
 \ifx #1\expandafter \@firstoftwo
 \else \expandafter \@secondoftwo
 \fi
}%
\providecommand \natexlab [1]{#1}%
\providecommand \enquote  [1]{``#1''}%
\providecommand \bibnamefont  [1]{#1}%
\providecommand \bibfnamefont [1]{#1}%
\providecommand \citenamefont [1]{#1}%
\providecommand \href@noop [0]{\@secondoftwo}%
\providecommand \href [0]{\begingroup \@sanitize@url \@href}%
\providecommand \@href[1]{\@@startlink{#1}\@@href}%
\providecommand \@@href[1]{\endgroup#1\@@endlink}%
\providecommand \@sanitize@url [0]{\catcode `\\12\catcode `\$12\catcode
  `\&12\catcode `\#12\catcode `\^12\catcode `\_12\catcode `\%12\relax}%
\providecommand \@@startlink[1]{}%
\providecommand \@@endlink[0]{}%
\providecommand \url  [0]{\begingroup\@sanitize@url \@url }%
\providecommand \@url [1]{\endgroup\@href {#1}{\urlprefix }}%
\providecommand \urlprefix  [0]{URL }%
\providecommand \Eprint [0]{\href }%
\providecommand \doibase [0]{http://dx.doi.org/}%
\providecommand \selectlanguage [0]{\@gobble}%
\providecommand \bibinfo  [0]{\@secondoftwo}%
\providecommand \bibfield  [0]{\@secondoftwo}%
\providecommand \translation [1]{[#1]}%
\providecommand \BibitemOpen [0]{}%
\providecommand \bibitemStop [0]{}%
\providecommand \bibitemNoStop [0]{.\EOS\space}%
\providecommand \EOS [0]{\spacefactor3000\relax}%
\providecommand \BibitemShut  [1]{\csname bibitem#1\endcsname}%
\let\auto@bib@innerbib\@empty
%</preamble>
\bibitem [{\citenamefont {Castro~Neto}\ \emph {et~al.}(2009)\citenamefont
  {Castro~Neto}, \citenamefont {Guinea}, \citenamefont {Peres}, \citenamefont
  {Novoselov},\ and\ \citenamefont {Geim}}]{25.2009Neto_electronic_properties}%
  \BibitemOpen
  \bibfield  {author} {\bibinfo {author} {\bibfnamefont {A.~H.}\ \bibnamefont
  {Castro~Neto}}, \bibinfo {author} {\bibfnamefont {F.}~\bibnamefont {Guinea}},
  \bibinfo {author} {\bibfnamefont {N.~M.~R.}\ \bibnamefont {Peres}}, \bibinfo
  {author} {\bibfnamefont {K.~S.}\ \bibnamefont {Novoselov}}, \ and\ \bibinfo
  {author} {\bibfnamefont {A.~K.}\ \bibnamefont {Geim}},\ }\href {\doibase
  10.1103/RevModPhys.81.109} {\bibfield  {journal} {\bibinfo  {journal} {Rev.
  Mod. Phys.}\ }\textbf {\bibinfo {volume} {81}},\ \bibinfo {pages} {109}
  (\bibinfo {year} {2009})}\BibitemShut {NoStop}%
\bibitem [{\citenamefont {Das~Sarma}\ \emph {et~al.}(2011)\citenamefont
  {Das~Sarma}, \citenamefont {Adam}, \citenamefont {Hwang},\ and\ \citenamefont
  {Rossi}}]{26.2011Sdas_etransport}%
  \BibitemOpen
  \bibfield  {author} {\bibinfo {author} {\bibfnamefont {S.}~\bibnamefont
  {Das~Sarma}}, \bibinfo {author} {\bibfnamefont {S.}~\bibnamefont {Adam}},
  \bibinfo {author} {\bibfnamefont {E.~H.}\ \bibnamefont {Hwang}}, \ and\
  \bibinfo {author} {\bibfnamefont {E.}~\bibnamefont {Rossi}},\ }\href
  {\doibase 10.1103/RevModPhys.83.407} {\bibfield  {journal} {\bibinfo
  {journal} {Rev. Mod. Phys.}\ }\textbf {\bibinfo {volume} {83}},\ \bibinfo
  {pages} {407} (\bibinfo {year} {2011})}\BibitemShut {NoStop}%
\bibitem [{\citenamefont {Dugaev}\ \emph {et~al.}(2011)\citenamefont {Dugaev},
  \citenamefont {Sherman},\ and\ \citenamefont {Barna\ifmmode~\acute{s}\else
  \'{s}\fi{}}}]{27.2011Dugaev_SOI}%
  \BibitemOpen
  \bibfield  {author} {\bibinfo {author} {\bibfnamefont {V.~K.}\ \bibnamefont
  {Dugaev}}, \bibinfo {author} {\bibfnamefont {E.~Y.}\ \bibnamefont {Sherman}},
  \ and\ \bibinfo {author} {\bibfnamefont {J.}~\bibnamefont
  {Barna\ifmmode~\acute{s}\else \'{s}\fi{}}},\ }\href {\doibase
  10.1103/PhysRevB.83.085306} {\bibfield  {journal} {\bibinfo  {journal} {Phys.
  Rev. B}\ }\textbf {\bibinfo {volume} {83}},\ \bibinfo {pages} {085306}
  (\bibinfo {year} {2011})}\BibitemShut {NoStop}%
\bibitem [{\citenamefont {Min}\ \emph {et~al.}(2006)\citenamefont {Min},
  \citenamefont {Hill}, \citenamefont {Sinitsyn}, \citenamefont {Sahu},
  \citenamefont {Kleinman},\ and\ \citenamefont
  {MacDonald}}]{28.2006Macdonald_RSOI}%
  \BibitemOpen
  \bibfield  {author} {\bibinfo {author} {\bibfnamefont {H.}~\bibnamefont
  {Min}}, \bibinfo {author} {\bibfnamefont {J.~E.}\ \bibnamefont {Hill}},
  \bibinfo {author} {\bibfnamefont {N.~A.}\ \bibnamefont {Sinitsyn}}, \bibinfo
  {author} {\bibfnamefont {B.~R.}\ \bibnamefont {Sahu}}, \bibinfo {author}
  {\bibfnamefont {L.}~\bibnamefont {Kleinman}}, \ and\ \bibinfo {author}
  {\bibfnamefont {A.~H.}\ \bibnamefont {MacDonald}},\ }\href {\doibase
  10.1103/PhysRevB.74.165310} {\bibfield  {journal} {\bibinfo  {journal} {Phys.
  Rev. B}\ }\textbf {\bibinfo {volume} {74}},\ \bibinfo {pages} {165310}
  (\bibinfo {year} {2006})}\BibitemShut {NoStop}%
\bibitem [{\citenamefont {Pesin}\ and\ \citenamefont
  {MacDonald}(2012)}]{29.2012Macdonald_spintronics}%
  \BibitemOpen
  \bibfield  {author} {\bibinfo {author} {\bibfnamefont {D.}~\bibnamefont
  {Pesin}}\ and\ \bibinfo {author} {\bibfnamefont {A.~H.}\ \bibnamefont
  {MacDonald}},\ }\href {\doibase 10.1038/nmat3305} {\bibfield  {journal}
  {\bibinfo  {journal} {Nature Materials}\ }\textbf {\bibinfo {volume} {11}},\
  \bibinfo {pages} {409} (\bibinfo {year} {2012})}\BibitemShut {NoStop}%
\bibitem [{\citenamefont {Drogeler}\ \emph {et~al.}(2016)\citenamefont
  {Drogeler}, \citenamefont {Franzen}, \citenamefont {Volmer}, \citenamefont
  {Pohlmann}, \citenamefont {Banszerus}, \citenamefont {Wolter}, \citenamefont
  {Watanabe}, \citenamefont {Taniguchi}, \citenamefont {Stampfer},\ and\
  \citenamefont {Beschoten}}]{41.2016.Drogler_12ns}%
  \BibitemOpen
  \bibfield  {author} {\bibinfo {author} {\bibfnamefont {M.}~\bibnamefont
  {Drogeler}}, \bibinfo {author} {\bibfnamefont {C.}~\bibnamefont {Franzen}},
  \bibinfo {author} {\bibfnamefont {F.}~\bibnamefont {Volmer}}, \bibinfo
  {author} {\bibfnamefont {T.}~\bibnamefont {Pohlmann}}, \bibinfo {author}
  {\bibfnamefont {L.}~\bibnamefont {Banszerus}}, \bibinfo {author}
  {\bibfnamefont {M.}~\bibnamefont {Wolter}}, \bibinfo {author} {\bibfnamefont
  {K.}~\bibnamefont {Watanabe}}, \bibinfo {author} {\bibfnamefont
  {T.}~\bibnamefont {Taniguchi}}, \bibinfo {author} {\bibfnamefont
  {C.}~\bibnamefont {Stampfer}}, \ and\ \bibinfo {author} {\bibfnamefont
  {B.}~\bibnamefont {Beschoten}},\ }\href {http://arxiv.org/abs/1602.02725}
  {\bibfield  {journal} {\bibinfo  {journal} {arXiv:1602.02725 [cond-mat]}\ }
  (\bibinfo {year} {2016})}\BibitemShut {NoStop}%
\bibitem [{\citenamefont {Roche}\ and\ \citenamefont
  {Valenzuela}(2014)}]{39.2014.Stephan_challenges}%
  \BibitemOpen
  \bibfield  {author} {\bibinfo {author} {\bibfnamefont {S.}~\bibnamefont
  {Roche}}\ and\ \bibinfo {author} {\bibfnamefont {S.~O.}\ \bibnamefont
  {Valenzuela}},\ }\href {http://stacks.iop.org/0022-3727/47/i=9/a=094011}
  {\bibfield  {journal} {\bibinfo  {journal} {Journal of Physics D: Applied
  Physics}\ }\textbf {\bibinfo {volume} {47}},\ \bibinfo {pages} {094011}
  (\bibinfo {year} {2014})}\BibitemShut {NoStop}%
\bibitem [{\citenamefont {Wang}\ \emph {et~al.}(2013)\citenamefont {Wang},
  \citenamefont {Meric}, \citenamefont {Huang}, \citenamefont {Gao},
  \citenamefont {Gao}, \citenamefont {Tran}, \citenamefont {Taniguchi},
  \citenamefont {Watanabe}, \citenamefont {Campos}, \citenamefont {Muller},
  \citenamefont {Guo}, \citenamefont {Kim}, \citenamefont {Hone}, \citenamefont
  {Shepard},\ and\ \citenamefont {Dean}}]{18.Dean_1Dcontacts}%
  \BibitemOpen
  \bibfield  {author} {\bibinfo {author} {\bibfnamefont {L.}~\bibnamefont
  {Wang}}, \bibinfo {author} {\bibfnamefont {I.}~\bibnamefont {Meric}},
  \bibinfo {author} {\bibfnamefont {P.~Y.}\ \bibnamefont {Huang}}, \bibinfo
  {author} {\bibfnamefont {Q.}~\bibnamefont {Gao}}, \bibinfo {author}
  {\bibfnamefont {Y.}~\bibnamefont {Gao}}, \bibinfo {author} {\bibfnamefont
  {H.}~\bibnamefont {Tran}}, \bibinfo {author} {\bibfnamefont {T.}~\bibnamefont
  {Taniguchi}}, \bibinfo {author} {\bibfnamefont {K.}~\bibnamefont {Watanabe}},
  \bibinfo {author} {\bibfnamefont {L.~M.}\ \bibnamefont {Campos}}, \bibinfo
  {author} {\bibfnamefont {D.~A.}\ \bibnamefont {Muller}}, \bibinfo {author}
  {\bibfnamefont {J.}~\bibnamefont {Guo}}, \bibinfo {author} {\bibfnamefont
  {P.}~\bibnamefont {Kim}}, \bibinfo {author} {\bibfnamefont {J.}~\bibnamefont
  {Hone}}, \bibinfo {author} {\bibfnamefont {K.~L.}\ \bibnamefont {Shepard}}, \
  and\ \bibinfo {author} {\bibfnamefont {C.~R.}\ \bibnamefont {Dean}},\ }\href
  {\doibase 10.1126/science.1244358} {\bibfield  {journal} {\bibinfo  {journal}
  {Science}\ }\textbf {\bibinfo {volume} {342}},\ \bibinfo {pages} {614}
  (\bibinfo {year} {2013})}\BibitemShut {NoStop}%
\bibitem [{\citenamefont {Zomer}\ \emph {et~al.}(2012)\citenamefont {Zomer},
  \citenamefont {GuimarÃ£es}, \citenamefont {Tombros},\ and\ \citenamefont {van
  Wees}}]{14.Paul_GrhBN}%
  \BibitemOpen
  \bibfield  {author} {\bibinfo {author} {\bibfnamefont {P.~J.}\ \bibnamefont
  {Zomer}}, \bibinfo {author} {\bibfnamefont {M.~H.~D.}\ \bibnamefont
  {Guimaraes}}, \bibinfo {author} {\bibfnamefont {N.}~\bibnamefont {Tombros}},
  \ and\ \bibinfo {author} {\bibfnamefont {B.~J.}\ \bibnamefont {van Wees}},\
  }\href {\doibase 10.1103/PhysRevB.86.161416} {\bibfield  {journal} {\bibinfo
  {journal} {Physical Review B}\ }\textbf {\bibinfo {volume} {86}},\ \bibinfo
  {pages} {161416} (\bibinfo {year} {2012})}\BibitemShut {NoStop}%
\bibitem [{\citenamefont {Van~Tuan}\ \emph {et~al.}(2016)\citenamefont
  {Van~Tuan}, \citenamefont {Ortmann}, \citenamefont {Cummings}, \citenamefont
  {Soriano},\ and\ \citenamefont {Roche}}]{37.2015.Dinh_spin_ehpuddles}%
  \BibitemOpen
  \bibfield  {author} {\bibinfo {author} {\bibfnamefont {D.}~\bibnamefont
  {Van~Tuan}}, \bibinfo {author} {\bibfnamefont {F.}~\bibnamefont {Ortmann}},
  \bibinfo {author} {\bibfnamefont {A.~W.}\ \bibnamefont {Cummings}}, \bibinfo
  {author} {\bibfnamefont {D.}~\bibnamefont {Soriano}}, \ and\ \bibinfo
  {author} {\bibfnamefont {S.}~\bibnamefont {Roche}},\ }\href {\doibase
  10.1038/srep21046} {\bibfield  {journal} {\bibinfo  {journal} {Scientific
  Reports}\ }\textbf {\bibinfo {volume} {6}},\ \bibinfo {pages} {21046}
  (\bibinfo {year} {2016})}\BibitemShut {NoStop}%
\bibitem [{\citenamefont {Kretinin}\ \emph {et~al.}(2014)\citenamefont
  {Kretinin}, \citenamefont {Cao}, \citenamefont {Tu}, \citenamefont {Yu},
  \citenamefont {Jalil}, \citenamefont {Novoselov}, \citenamefont {Haigh},
  \citenamefont {Gholinia}, \citenamefont {Mishchenko}, \citenamefont {Lozada},
  \citenamefont {Georgiou}, \citenamefont {Woods}, \citenamefont {Withers},
  \citenamefont {Blake}, \citenamefont {Eda}, \citenamefont {Wirsig},
  \citenamefont {Hucho}, \citenamefont {Watanabe}, \citenamefont {Taniguchi},
  \citenamefont {Geim},\ and\ \citenamefont
  {Gorbachev}}]{10.Kretinin_hBNGrMoS2_NL}%
  \BibitemOpen
  \bibfield  {author} {\bibinfo {author} {\bibfnamefont {A.~V.}\ \bibnamefont
  {Kretinin}}, \bibinfo {author} {\bibfnamefont {Y.}~\bibnamefont {Cao}},
  \bibinfo {author} {\bibfnamefont {J.~S.}\ \bibnamefont {Tu}}, \bibinfo
  {author} {\bibfnamefont {G.~L.}\ \bibnamefont {Yu}}, \bibinfo {author}
  {\bibfnamefont {R.}~\bibnamefont {Jalil}}, \bibinfo {author} {\bibfnamefont
  {K.~S.}\ \bibnamefont {Novoselov}}, \bibinfo {author} {\bibfnamefont {S.~J.}\
  \bibnamefont {Haigh}}, \bibinfo {author} {\bibfnamefont {A.}~\bibnamefont
  {Gholinia}}, \bibinfo {author} {\bibfnamefont {A.}~\bibnamefont
  {Mishchenko}}, \bibinfo {author} {\bibfnamefont {M.}~\bibnamefont {Lozada}},
  \bibinfo {author} {\bibfnamefont {T.}~\bibnamefont {Georgiou}}, \bibinfo
  {author} {\bibfnamefont {C.~R.}\ \bibnamefont {Woods}}, \bibinfo {author}
  {\bibfnamefont {F.}~\bibnamefont {Withers}}, \bibinfo {author} {\bibfnamefont
  {P.}~\bibnamefont {Blake}}, \bibinfo {author} {\bibfnamefont
  {G.}~\bibnamefont {Eda}}, \bibinfo {author} {\bibfnamefont {A.}~\bibnamefont
  {Wirsig}}, \bibinfo {author} {\bibfnamefont {C.}~\bibnamefont {Hucho}},
  \bibinfo {author} {\bibfnamefont {K.}~\bibnamefont {Watanabe}}, \bibinfo
  {author} {\bibfnamefont {T.}~\bibnamefont {Taniguchi}}, \bibinfo {author}
  {\bibfnamefont {A.~K.}\ \bibnamefont {Geim}}, \ and\ \bibinfo {author}
  {\bibfnamefont {R.~V.}\ \bibnamefont {Gorbachev}},\ }\href {\doibase
  10.1021/nl5006542} {\bibfield  {journal} {\bibinfo  {journal} {Nano Letters}\
  }\textbf {\bibinfo {volume} {14}},\ \bibinfo {pages} {3270} (\bibinfo {year}
  {2014})}\BibitemShut {NoStop}%
\bibitem [{\citenamefont {Volmer}\ \emph {et~al.}(2014)\citenamefont {Volmer},
  \citenamefont {Dr\"ogeler}, \citenamefont {Maynicke}, \citenamefont {von~den
  Driesch}, \citenamefont {Boschen}, \citenamefont {G\"untherodt},
  \citenamefont {Stampfer},\ and\ \citenamefont
  {Beschoten}}]{30.2014Aachen_contacts_oxidation}%
  \BibitemOpen
  \bibfield  {author} {\bibinfo {author} {\bibfnamefont {F.}~\bibnamefont
  {Volmer}}, \bibinfo {author} {\bibfnamefont {M.}~\bibnamefont {Dr\"ogeler}},
  \bibinfo {author} {\bibfnamefont {E.}~\bibnamefont {Maynicke}}, \bibinfo
  {author} {\bibfnamefont {N.}~\bibnamefont {von~den Driesch}}, \bibinfo
  {author} {\bibfnamefont {M.~L.}\ \bibnamefont {Boschen}}, \bibinfo {author}
  {\bibfnamefont {G.}~\bibnamefont {G\"untherodt}}, \bibinfo {author}
  {\bibfnamefont {C.}~\bibnamefont {Stampfer}}, \ and\ \bibinfo {author}
  {\bibfnamefont {B.}~\bibnamefont {Beschoten}},\ }\href {\doibase
  10.1103/PhysRevB.90.165403} {\bibfield  {journal} {\bibinfo  {journal} {Phys.
  Rev. B}\ }\textbf {\bibinfo {volume} {90}},\ \bibinfo {pages} {165403}
  (\bibinfo {year} {2014})}\BibitemShut {NoStop}%
\bibitem [{\citenamefont {Lee}\ \emph {et~al.}(2011)\citenamefont {Lee},
  \citenamefont {Yu}, \citenamefont {Lee}, \citenamefont {Dean}, \citenamefont
  {Shepard}, \citenamefont {Kim},\ and\ \citenamefont
  {Hone}}]{38.2011.Lee_hBNtunnel}%
  \BibitemOpen
  \bibfield  {author} {\bibinfo {author} {\bibfnamefont {G.-H.}\ \bibnamefont
  {Lee}}, \bibinfo {author} {\bibfnamefont {Y.-J.}\ \bibnamefont {Yu}},
  \bibinfo {author} {\bibfnamefont {C.}~\bibnamefont {Lee}}, \bibinfo {author}
  {\bibfnamefont {C.}~\bibnamefont {Dean}}, \bibinfo {author} {\bibfnamefont
  {K.~L.}\ \bibnamefont {Shepard}}, \bibinfo {author} {\bibfnamefont
  {P.}~\bibnamefont {Kim}}, \ and\ \bibinfo {author} {\bibfnamefont
  {J.}~\bibnamefont {Hone}},\ }\href {\doibase 10.1063/1.3662043} {\bibfield
  {journal} {\bibinfo  {journal} {Applied Physics Letters}\ }\textbf {\bibinfo
  {volume} {99}},\ \bibinfo {pages} {243114} (\bibinfo {year}
  {2011})}\BibitemShut {NoStop}%
\bibitem [{\citenamefont {Britnell}\ \emph {et~al.}(2012)\citenamefont
  {Britnell}, \citenamefont {Gorbachev}, \citenamefont {Jalil}, \citenamefont
  {Belle}, \citenamefont {Schedin}, \citenamefont {Katsnelson}, \citenamefont
  {Eaves}, \citenamefont {Morozov}, \citenamefont {Mayorov}, \citenamefont
  {Peres}, \citenamefont {Neto}, \citenamefont {Leist}, \citenamefont {Geim},
  \citenamefont {Ponomarenko},\ and\ \citenamefont
  {Novoselov}}]{6.Britnell_hBNtb}%
  \BibitemOpen
  \bibfield  {author} {\bibinfo {author} {\bibfnamefont {L.}~\bibnamefont
  {Britnell}}, \bibinfo {author} {\bibfnamefont {R.~V.}\ \bibnamefont
  {Gorbachev}}, \bibinfo {author} {\bibfnamefont {R.}~\bibnamefont {Jalil}},
  \bibinfo {author} {\bibfnamefont {B.~D.}\ \bibnamefont {Belle}}, \bibinfo
  {author} {\bibfnamefont {F.}~\bibnamefont {Schedin}}, \bibinfo {author}
  {\bibfnamefont {M.~I.}\ \bibnamefont {Katsnelson}}, \bibinfo {author}
  {\bibfnamefont {L.}~\bibnamefont {Eaves}}, \bibinfo {author} {\bibfnamefont
  {S.~V.}\ \bibnamefont {Morozov}}, \bibinfo {author} {\bibfnamefont {A.~S.}\
  \bibnamefont {Mayorov}}, \bibinfo {author} {\bibfnamefont {N.~M.~R.}\
  \bibnamefont {Peres}}, \bibinfo {author} {\bibfnamefont {A.~H.~C.}\
  \bibnamefont {Neto}}, \bibinfo {author} {\bibfnamefont {J.}~\bibnamefont
  {Leist}}, \bibinfo {author} {\bibfnamefont {A.~K.}\ \bibnamefont {Geim}},
  \bibinfo {author} {\bibfnamefont {L.~A.}\ \bibnamefont {Ponomarenko}}, \ and\
  \bibinfo {author} {\bibfnamefont {K.~S.}\ \bibnamefont {Novoselov}},\ }\href
  {\doibase 10.1021/nl3002205} {\bibfield  {journal} {\bibinfo  {journal} {Nano
  Letters}\ }\textbf {\bibinfo {volume} {12}},\ \bibinfo {pages} {1707}
  (\bibinfo {year} {2012})}\BibitemShut {NoStop}%
\bibitem [{\citenamefont {Wu}\ \emph {et~al.}(2014)\citenamefont {Wu},
  \citenamefont {Shen}, \citenamefont {Bai}, \citenamefont {Zeng},
  \citenamefont {Yang}, \citenamefont {Huang},\ and\ \citenamefont
  {Feng}}]{7.Theory_hBNtb}%
  \BibitemOpen
  \bibfield  {author} {\bibinfo {author} {\bibfnamefont {Q.}~\bibnamefont
  {Wu}}, \bibinfo {author} {\bibfnamefont {L.}~\bibnamefont {Shen}}, \bibinfo
  {author} {\bibfnamefont {Z.}~\bibnamefont {Bai}}, \bibinfo {author}
  {\bibfnamefont {M.}~\bibnamefont {Zeng}}, \bibinfo {author} {\bibfnamefont
  {M.}~\bibnamefont {Yang}}, \bibinfo {author} {\bibfnamefont {Z.}~\bibnamefont
  {Huang}}, \ and\ \bibinfo {author} {\bibfnamefont {Y.~P.}\ \bibnamefont
  {Feng}},\ }\href {\doibase 10.1103/PhysRevApplied.2.044008} {\bibfield
  {journal} {\bibinfo  {journal} {Phys. Rev. Applied}\ }\textbf {\bibinfo
  {volume} {2}},\ \bibinfo {pages} {044008} (\bibinfo {year}
  {2014})}\BibitemShut {NoStop}%
\bibitem [{\citenamefont {Yamaguchi}\ \emph {et~al.}(2013)\citenamefont
  {Yamaguchi}, \citenamefont {Inoue}, \citenamefont {Masubuchi}, \citenamefont
  {Morikawa}, \citenamefont {Onuki}, \citenamefont {Watanabe}, \citenamefont
  {Taniguchi}, \citenamefont {Moriya},\ and\ \citenamefont
  {Machida}}]{2.Yamaguchi_exhBNtb}%
  \BibitemOpen
  \bibfield  {author} {\bibinfo {author} {\bibfnamefont {T.}~\bibnamefont
  {Yamaguchi}}, \bibinfo {author} {\bibfnamefont {Y.}~\bibnamefont {Inoue}},
  \bibinfo {author} {\bibfnamefont {S.}~\bibnamefont {Masubuchi}}, \bibinfo
  {author} {\bibfnamefont {S.}~\bibnamefont {Morikawa}}, \bibinfo {author}
  {\bibfnamefont {M.}~\bibnamefont {Onuki}}, \bibinfo {author} {\bibfnamefont
  {K.}~\bibnamefont {Watanabe}}, \bibinfo {author} {\bibfnamefont
  {T.}~\bibnamefont {Taniguchi}}, \bibinfo {author} {\bibfnamefont
  {R.}~\bibnamefont {Moriya}}, \ and\ \bibinfo {author} {\bibfnamefont
  {T.}~\bibnamefont {Machida}},\ }\href
  {http://stacks.iop.org/1882-0786/6/i=7/a=073001} {\bibfield  {journal}
  {\bibinfo  {journal} {Applied Physics Express}\ }\textbf {\bibinfo {volume}
  {6}},\ \bibinfo {pages} {073001} (\bibinfo {year} {2013})}\BibitemShut
  {NoStop}%
\bibitem [{\citenamefont {Kamalakar}\ \emph {et~al.}(2014)\citenamefont
  {Kamalakar}, \citenamefont {Dankert}, \citenamefont {Bergsten}, \citenamefont
  {Ive},\ and\ \citenamefont {Dash}}]{3.Kamalakar_CVDhBNtb}%
  \BibitemOpen
  \bibfield  {author} {\bibinfo {author} {\bibfnamefont {M.~V.}\ \bibnamefont
  {Kamalakar}}, \bibinfo {author} {\bibfnamefont {A.}~\bibnamefont {Dankert}},
  \bibinfo {author} {\bibfnamefont {J.}~\bibnamefont {Bergsten}}, \bibinfo
  {author} {\bibfnamefont {T.}~\bibnamefont {Ive}}, \ and\ \bibinfo {author}
  {\bibfnamefont {S.~P.}\ \bibnamefont {Dash}},\ }\href {\doibase
  10.1038/srep06146} {\bibfield  {journal} {\bibinfo  {journal} {Scientific
  Reports}\ }\textbf {\bibinfo {volume} {4}},\ \bibinfo {pages} {6146}
  (\bibinfo {year} {2014})}\BibitemShut {NoStop}%
\bibitem [{\citenamefont {Fu}\ \emph {et~al.}(2014)\citenamefont {Fu},
  \citenamefont {Makk}, \citenamefont {Maurand}, \citenamefont {BrÃ¤uninger},\
  and\ \citenamefont {SchÃ¶nenberger}}]{4.Basel_CVDhBNtb}%
  \BibitemOpen
  \bibfield  {author} {\bibinfo {author} {\bibfnamefont {W.}~\bibnamefont
  {Fu}}, \bibinfo {author} {\bibfnamefont {P.}~\bibnamefont {Makk}}, \bibinfo
  {author} {\bibfnamefont {R.}~\bibnamefont {Maurand}}, \bibinfo {author}
  {\bibfnamefont {M.}~\bibnamefont {BrÃ¤uninger}}, \ and\ \bibinfo {author}
  {\bibfnamefont {C.}~\bibnamefont {SchÃ¶nenberger}},\ }\href
  {http://scitation.aip.org/content/aip/journal/jap/116/7/10.1063/1.4893578}
  {\bibfield  {journal} {\bibinfo  {journal} {Journal of Applied Physics}\
  }\textbf {\bibinfo {volume} {116}},\ \bibinfo {eid} {074306} (\bibinfo {year}
  {2014})}\BibitemShut {NoStop}%
\bibitem [{\citenamefont {Zomer}\ \emph {et~al.}(2014)\citenamefont {Zomer},
  \citenamefont {GuimarÃ£es}, \citenamefont {Brant}, \citenamefont {Tombros},\
  and\ \citenamefont {Wees}}]{19.Paul_fastpickup}%
  \BibitemOpen
  \bibfield  {author} {\bibinfo {author} {\bibfnamefont {P.~J.}\ \bibnamefont
  {Zomer}}, \bibinfo {author} {\bibfnamefont {M.~H.~D.}\ \bibnamefont
  {GuimarÃ£es}}, \bibinfo {author} {\bibfnamefont {J.~C.}\ \bibnamefont
  {Brant}}, \bibinfo {author} {\bibfnamefont {N.}~\bibnamefont {Tombros}}, \
  and\ \bibinfo {author} {\bibfnamefont {B.~J.~v.}\ \bibnamefont {Wees}},\
  }\href {\doibase 10.1063/1.4886096} {\bibfield  {journal} {\bibinfo
  {journal} {Applied Physics Letters}\ }\textbf {\bibinfo {volume} {105}},\
  \bibinfo {pages} {013101} (\bibinfo {year} {2014})}\BibitemShut {NoStop}%
\bibitem [{\citenamefont {GuimarÃ£es}\ \emph {et~al.}(2014)\citenamefont
  {GuimarÃ£es}, \citenamefont {Zomer}, \citenamefont {Ingla-AynÃ©s},
  \citenamefont {Brant}, \citenamefont {Tombros},\ and\ \citenamefont {van
  Wees}}]{17.Marcos_hBNencapsulation}%
  \BibitemOpen
  \bibfield  {author} {\bibinfo {author} {\bibfnamefont {M.H.D.}~\bibnamefont
  {Guimaraes}}, \bibinfo {author} {\bibfnamefont {P.J.}~\bibnamefont {Zomer}},
  \bibinfo {author} {\bibfnamefont {J.}~\bibnamefont {Ingla-Aynes}}, \bibinfo
  {author} {\bibfnamefont {J.C.}~\bibnamefont {Brant}}, \bibinfo {author}
  {\bibfnamefont {N.}~\bibnamefont {Tombros}}, \ and\ \bibinfo {author}
  {\bibfnamefont {B.J.}~\bibnamefont {van Wees}},\ }\href {\doibase
  10.1103/PhysRevLett.113.086602} {\bibfield  {journal} {\bibinfo  {journal}
  {Physical Review Letters}\ }\textbf {\bibinfo {volume} {113}},\ \bibinfo
  {pages} {086602} (\bibinfo {year} {2014})}\BibitemShut {NoStop}%
\bibitem [{\citenamefont {Ingla-Ayn\'es}\ \emph {et~al.}(2015)\citenamefont
  {Ingla-Ayn\'es}, \citenamefont {Guimar\~aes}, \citenamefont {Meijerink},
  \citenamefont {Zomer},\ and\ \citenamefont {van Wees}}]{16.Pep_24um}%
  \BibitemOpen
  \bibfield  {author} {\bibinfo {author} {\bibfnamefont {J.}~\bibnamefont
  {Ingla-Ayn\'es}}, \bibinfo {author} {\bibfnamefont {M.~H.~D.}\ \bibnamefont
  {Guimar\~aes}}, \bibinfo {author} {\bibfnamefont {R.~J.}\ \bibnamefont
  {Meijerink}}, \bibinfo {author} {\bibfnamefont {P.~J.}\ \bibnamefont
  {Zomer}}, \ and\ \bibinfo {author} {\bibfnamefont {B.~J.}\ \bibnamefont {van
  Wees}},\ }\href {\doibase 10.1103/PhysRevB.92.201410} {\bibfield  {journal}
  {\bibinfo  {journal} {Phys. Rev. B}\ }\textbf {\bibinfo {volume} {92}},\
  \bibinfo {pages} {201410} (\bibinfo {year} {2015})}\BibitemShut {NoStop}%
\bibitem [{\citenamefont {GuimarÃ£es}\ \emph {et~al.}(2012)\citenamefont
  {GuimarÃ£es}, \citenamefont {Veligura}, \citenamefont {Zomer}, \citenamefont
  {Maassen}, \citenamefont {Vera-Marun}, \citenamefont {Tombros},\ and\
  \citenamefont {van Wees}}]{15.Marcos_GrSuspended}%
  \BibitemOpen
  \bibfield  {author} {\bibinfo {author} {\bibfnamefont {M.~H.~D.}\
  \bibnamefont {GuimarÃ£es}}, \bibinfo {author} {\bibfnamefont
  {A.}~\bibnamefont {Veligura}}, \bibinfo {author} {\bibfnamefont {P.~J.}\
  \bibnamefont {Zomer}}, \bibinfo {author} {\bibfnamefont {T.}~\bibnamefont
  {Maassen}}, \bibinfo {author} {\bibfnamefont {I.~J.}\ \bibnamefont
  {Vera-Marun}}, \bibinfo {author} {\bibfnamefont {N.}~\bibnamefont {Tombros}},
  \ and\ \bibinfo {author} {\bibfnamefont {B.~J.}\ \bibnamefont {van Wees}},\
  }\href {\doibase 10.1021/nl301050a} {\bibfield  {journal} {\bibinfo
  {journal} {Nano Letters}\ }\textbf {\bibinfo {volume} {12}},\ \bibinfo
  {pages} {3512} (\bibinfo {year} {2012})}\BibitemShut {NoStop}%
\bibitem [{\citenamefont {Blake}\ \emph {et~al.}(2007)\citenamefont {Blake},
  \citenamefont {Hill}, \citenamefont {Castro~Neto}, \citenamefont {Novoselov},
  \citenamefont {Jiang}, \citenamefont {Yang}, \citenamefont {Booth},\ and\
  \citenamefont {Geim}}]{1b.Hunting_graphene}%
  \BibitemOpen
  \bibfield  {author} {\bibinfo {author} {\bibfnamefont {P.}~\bibnamefont
  {Blake}}, \bibinfo {author} {\bibfnamefont {E.~W.}\ \bibnamefont {Hill}},
  \bibinfo {author} {\bibfnamefont {A.~H.}\ \bibnamefont {Castro~Neto}},
  \bibinfo {author} {\bibfnamefont {K.~S.}\ \bibnamefont {Novoselov}}, \bibinfo
  {author} {\bibfnamefont {D.}~\bibnamefont {Jiang}}, \bibinfo {author}
  {\bibfnamefont {R.}~\bibnamefont {Yang}}, \bibinfo {author} {\bibfnamefont
  {T.~J.}\ \bibnamefont {Booth}}, \ and\ \bibinfo {author} {\bibfnamefont
  {A.~K.}\ \bibnamefont {Geim}},\ }\href
  {http://scitation.aip.org/content/aip/journal/apl/91/6/10.1063/1.2768624}
  {\bibfield  {journal} {\bibinfo  {journal} {Applied Physics Letters}\
  }\textbf {\bibinfo {volume} {91}},\ \bibinfo {eid} {063124} (\bibinfo {year}
  {2007})}\BibitemShut {NoStop}%
\bibitem [{\citenamefont {Gorbachev}\ \emph {et~al.}(2011)\citenamefont
  {Gorbachev}, \citenamefont {Riaz}, \citenamefont {Nair}, \citenamefont
  {Jalil}, \citenamefont {Britnell}, \citenamefont {Belle}, \citenamefont
  {Hill}, \citenamefont {Novoselov}, \citenamefont {Watanabe}, \citenamefont
  {Taniguchi}, \citenamefont {Geim},\ and\ \citenamefont
  {Blake}}]{1_Hunting_hBN}%
  \BibitemOpen
  \bibfield  {author} {\bibinfo {author} {\bibfnamefont {R.~V.}\ \bibnamefont
  {Gorbachev}}, \bibinfo {author} {\bibfnamefont {I.}~\bibnamefont {Riaz}},
  \bibinfo {author} {\bibfnamefont {R.~R.}\ \bibnamefont {Nair}}, \bibinfo
  {author} {\bibfnamefont {R.}~\bibnamefont {Jalil}}, \bibinfo {author}
  {\bibfnamefont {L.}~\bibnamefont {Britnell}}, \bibinfo {author}
  {\bibfnamefont {B.~D.}\ \bibnamefont {Belle}}, \bibinfo {author}
  {\bibfnamefont {E.~W.}\ \bibnamefont {Hill}}, \bibinfo {author}
  {\bibfnamefont {K.~S.}\ \bibnamefont {Novoselov}}, \bibinfo {author}
  {\bibfnamefont {K.}~\bibnamefont {Watanabe}}, \bibinfo {author}
  {\bibfnamefont {T.}~\bibnamefont {Taniguchi}}, \bibinfo {author}
  {\bibfnamefont {A.~K.}\ \bibnamefont {Geim}}, \ and\ \bibinfo {author}
  {\bibfnamefont {P.}~\bibnamefont {Blake}},\ }\href {\doibase
  10.1002/smll.201001628} {\bibfield  {journal} {\bibinfo  {journal} {Small}\
  }\textbf {\bibinfo {volume} {7}},\ \bibinfo {pages} {465} (\bibinfo {year}
  {2011})}\BibitemShut {NoStop}%
\bibitem [{\citenamefont {Tombros}\ \emph {et~al.}(2007)\citenamefont
  {Tombros}, \citenamefont {Jozsa}, \citenamefont {Popinciuc}, \citenamefont
  {Jonkman},\ and\ \citenamefont {van Wees}}]{23.Nikos_2007}%
  \BibitemOpen
  \bibfield  {author} {\bibinfo {author} {\bibfnamefont {N.}~\bibnamefont
  {Tombros}}, \bibinfo {author} {\bibfnamefont {C.}~\bibnamefont {Jozsa}},
  \bibinfo {author} {\bibfnamefont {M.}~\bibnamefont {Popinciuc}}, \bibinfo
  {author} {\bibfnamefont {H.~T.}\ \bibnamefont {Jonkman}}, \ and\ \bibinfo
  {author} {\bibfnamefont {B.~J.}\ \bibnamefont {van Wees}},\ }\href {\doibase
  10.1038/nature06037} {\bibfield  {journal} {\bibinfo  {journal} {Nature}\
  }\textbf {\bibinfo {volume} {448}},\ \bibinfo {pages} {571} (\bibinfo {year}
  {2007})}\BibitemShut {NoStop}%
\bibitem [{\citenamefont {Morozov}\ \emph {et~al.}(2008)\citenamefont
  {Morozov}, \citenamefont {Novoselov}, \citenamefont {Katsnelson},
  \citenamefont {Schedin}, \citenamefont {Elias}, \citenamefont {Jaszczak},\
  and\ \citenamefont {Geim}}]{14b.Morozov_mobility}%
  \BibitemOpen
  \bibfield  {author} {\bibinfo {author} {\bibfnamefont {S.~V.}\ \bibnamefont
  {Morozov}}, \bibinfo {author} {\bibfnamefont {K.~S.}\ \bibnamefont
  {Novoselov}}, \bibinfo {author} {\bibfnamefont {M.~I.}\ \bibnamefont
  {Katsnelson}}, \bibinfo {author} {\bibfnamefont {F.}~\bibnamefont {Schedin}},
  \bibinfo {author} {\bibfnamefont {D.~C.}\ \bibnamefont {Elias}}, \bibinfo
  {author} {\bibfnamefont {J.~A.}\ \bibnamefont {Jaszczak}}, \ and\ \bibinfo
  {author} {\bibfnamefont {A.~K.}\ \bibnamefont {Geim}},\ }\href {\doibase
  10.1103/PhysRevLett.100.016602} {\bibfield  {journal} {\bibinfo  {journal}
  {Phys. Rev. Lett.}\ }\textbf {\bibinfo {volume} {100}},\ \bibinfo {pages}
  {016602} (\bibinfo {year} {2008})}\BibitemShut {NoStop}%
\bibitem [{\citenamefont {Zomer}\ \emph {et~al.}(2011)\citenamefont {Zomer},
  \citenamefont {Dash}, \citenamefont {Tombros},\ and\ \citenamefont {van
  Wees}}]{14c.Paul_mobility}%
  \BibitemOpen
  \bibfield  {author} {\bibinfo {author} {\bibfnamefont {P.~J.}\ \bibnamefont
  {Zomer}}, \bibinfo {author} {\bibfnamefont {S.~P.}\ \bibnamefont {Dash}},
  \bibinfo {author} {\bibfnamefont {N.}~\bibnamefont {Tombros}}, \ and\
  \bibinfo {author} {\bibfnamefont {B.~J.}\ \bibnamefont {van Wees}},\ }\href
  {http://scitation.aip.org/content/aip/journal/apl/99/23/10.1063/1.3665405}
  {\bibfield  {journal} {\bibinfo  {journal} {Applied Physics Letters}\
  }\textbf {\bibinfo {volume} {99}},\ \bibinfo {eid} {232104} (\bibinfo {year}
  {2011})}\BibitemShut {NoStop}%
\bibitem [{\citenamefont {Maassen}\ \emph {et~al.}(2012)\citenamefont
  {Maassen}, \citenamefont {Vera-Marun}, \citenamefont {Guimar\~aes},\ and\
  \citenamefont {van Wees}}]{34.2012Thomas_mismatch}%
  \BibitemOpen
  \bibfield  {author} {\bibinfo {author} {\bibfnamefont {T.}~\bibnamefont
  {Maassen}}, \bibinfo {author} {\bibfnamefont {I.~J.}\ \bibnamefont
  {Vera-Marun}}, \bibinfo {author} {\bibfnamefont {M.~H.~D.}\ \bibnamefont
  {Guimar\~aes}}, \ and\ \bibinfo {author} {\bibfnamefont {B.~J.}\ \bibnamefont
  {van Wees}},\ }\href {\doibase 10.1103/PhysRevB.86.235408} {\bibfield
  {journal} {\bibinfo  {journal} {Phys. Rev. B}\ }\textbf {\bibinfo {volume}
  {86}},\ \bibinfo {pages} {235408} (\bibinfo {year} {2012})}\BibitemShut
  {NoStop}%
\bibitem [{\citenamefont {Han}\ \emph {et~al.}(2010)\citenamefont {Han},
  \citenamefont {Pi}, \citenamefont {McCreary}, \citenamefont {Li},
  \citenamefont {Wong}, \citenamefont {Swartz},\ and\ \citenamefont
  {Kawakami}}]{31.2010Kawakami_3tunnelbarriers}%
  \BibitemOpen
  \bibfield  {author} {\bibinfo {author} {\bibfnamefont {W.}~\bibnamefont
  {Han}}, \bibinfo {author} {\bibfnamefont {K.}~\bibnamefont {Pi}}, \bibinfo
  {author} {\bibfnamefont {K.~M.}\ \bibnamefont {McCreary}}, \bibinfo {author}
  {\bibfnamefont {Y.}~\bibnamefont {Li}}, \bibinfo {author} {\bibfnamefont
  {J.~J.~I.}\ \bibnamefont {Wong}}, \bibinfo {author} {\bibfnamefont {A.~G.}\
  \bibnamefont {Swartz}}, \ and\ \bibinfo {author} {\bibfnamefont {R.~K.}\
  \bibnamefont {Kawakami}},\ }\href {\doibase 10.1103/PhysRevLett.105.167202}
  {\bibfield  {journal} {\bibinfo  {journal} {Phys. Rev. Lett.}\ }\textbf
  {\bibinfo {volume} {105}},\ \bibinfo {pages} {167202} (\bibinfo {year}
  {2010})}\BibitemShut {NoStop}%
\bibitem [{\citenamefont {Huang}\ \emph {et~al.}(2005)\citenamefont {Huang},
  \citenamefont {Lo}, \citenamefont {Yao}, \citenamefont {Hsieh},\ and\
  \citenamefont {Huang}}]{40.2005.Huang_SV_pn_transistor}%
  \BibitemOpen
  \bibfield  {author} {\bibinfo {author} {\bibfnamefont {Y.~W.}\ \bibnamefont
  {Huang}}, \bibinfo {author} {\bibfnamefont {C.~K.}\ \bibnamefont {Lo}},
  \bibinfo {author} {\bibfnamefont {Y.~D.}\ \bibnamefont {Yao}}, \bibinfo
  {author} {\bibfnamefont {L.~C.}\ \bibnamefont {Hsieh}}, \ and\ \bibinfo
  {author} {\bibfnamefont {J.~H.}\ \bibnamefont {Huang}},\ }\href@noop {}
  {\bibfield  {journal} {\bibinfo  {journal} {Journal of Applied Physics}\
  }\textbf {\bibinfo {volume} {97}},\ \bibinfo {eid} {10} (\bibinfo {year}
  {2005})}\BibitemShut {NoStop}%
\bibitem [{\citenamefont {Zhang}\ \emph {et~al.}(2009)\citenamefont {Zhang},
  \citenamefont {Tang}, \citenamefont {Girit}, \citenamefont {Hao},
  \citenamefont {Martin}, \citenamefont {Zettl}, \citenamefont {Crommie},
  \citenamefont {Shen},\ and\ \citenamefont
  {Wang}}]{35.2009Zhang_VbgVtgelectricfield}%
  \BibitemOpen
  \bibfield  {author} {\bibinfo {author} {\bibfnamefont {Y.}~\bibnamefont
  {Zhang}}, \bibinfo {author} {\bibfnamefont {T.-T.}\ \bibnamefont {Tang}},
  \bibinfo {author} {\bibfnamefont {C.}~\bibnamefont {Girit}}, \bibinfo
  {author} {\bibfnamefont {Z.}~\bibnamefont {Hao}}, \bibinfo {author}
  {\bibfnamefont {M.~C.}\ \bibnamefont {Martin}}, \bibinfo {author}
  {\bibfnamefont {A.}~\bibnamefont {Zettl}}, \bibinfo {author} {\bibfnamefont
  {M.~F.}\ \bibnamefont {Crommie}}, \bibinfo {author} {\bibfnamefont {Y.~R.}\
  \bibnamefont {Shen}}, \ and\ \bibinfo {author} {\bibfnamefont
  {F.}~\bibnamefont {Wang}},\ }\href {\doibase 10.1038/nature08105} {\bibfield
  {journal} {\bibinfo  {journal} {Nature}\ }\textbf {\bibinfo {volume} {459}},\
  \bibinfo {pages} {820} (\bibinfo {year} {2009})}\BibitemShut {NoStop}%
\bibitem [{\citenamefont {Williams}\ \emph {et~al.}(2007)\citenamefont
  {Williams}, \citenamefont {DiCarlo},\ and\ \citenamefont
  {Marcus}}]{36.2007.Williams_pnjunctions}%
  \BibitemOpen
  \bibfield  {author} {\bibinfo {author} {\bibfnamefont {J.~R.}\ \bibnamefont
  {Williams}}, \bibinfo {author} {\bibfnamefont {L.}~\bibnamefont {DiCarlo}}, \
  and\ \bibinfo {author} {\bibfnamefont {C.~M.}\ \bibnamefont {Marcus}},\
  }\href {\doibase 10.1126/science.1144657} {\bibfield  {journal} {\bibinfo
  {journal} {Science}\ }\textbf {\bibinfo {volume} {317}},\ \bibinfo {pages}
  {638} (\bibinfo {year} {2007})}\BibitemShut {NoStop}%
\bibitem [{\citenamefont {Han}\ and\ \citenamefont
  {Kawakami}(2011)}]{21.Kawakami_1L2L}%
  \BibitemOpen
  \bibfield  {author} {\bibinfo {author} {\bibfnamefont {W.}~\bibnamefont
  {Han}}\ and\ \bibinfo {author} {\bibfnamefont {R.~K.}\ \bibnamefont
  {Kawakami}},\ }\href {\doibase 10.1103/PhysRevLett.107.047207} {\bibfield
  {journal} {\bibinfo  {journal} {Physical Review Letters}\ }\textbf {\bibinfo
  {volume} {107}},\ \bibinfo {pages} {047207} (\bibinfo {year}
  {2011})}\BibitemShut {NoStop}%
\end{thebibliography}

% \begin{comment}


% \end{comment}

\end{document}